\documentclass[aps,pra,reprint,showpacs,groupedaddress]{revtex4-1} 

\usepackage{hyperref}  
\hypersetup{
pdfauthor={Otim Odong},
pdftitle={Heterodimer of two distinguishable atoms in a one-dimensional optical lattice}, colorlinks=true,urlcolor=blue,linkcolor=blue, bookmarks=true, pdftoolbar=true, filecolor=red, citecolor=blue}
\usepackage{graphicx}  
\usepackage{dcolumn}   
\usepackage{bm}        
\usepackage{amssymb}   
\usepackage{amsmath}   
\usepackage{verbatim}  
\usepackage{epstopdf}
\usepackage{color}
\newcommand{\ket}[1]{\ensuremath{\left| #1 \right\rangle}}

\newcommand{\mele}[3]{\ensuremath{\left\langle #1 \right|#2\left| #3
\right\rangle}}

\newcommand{\abs}[1]{\ensuremath{\left| #1 \right|}}

\newcommand{\beq}{\begin{equation}}
\newcommand{\eeq}{\end{equation}}
\newcommand{\bea}{\begin{eqnarray}}
\newcommand{\eea}{\end{eqnarray}}

\newcommand{\eq}[1]{{(\ref{#1})}}
\newcommand{\commentout}[1]{{}}

\newcommand{\half}{{\hbox{$\frac{1}{2}$}}}
\newcommand{\Kappa}{{\cal K}}
\definecolor{red}{rgb}{1,0,0}

\begin{document}

\title{Heterodimer of two distinguishable atoms in a one-dimensional optical lattice}

\author{Otim Odong}
\affiliation{Department of Physics, University of Connecticut, Storrs, Connecticut 06269-3046}
\author{Jerome C. Sanders}
\affiliation{Department of Physics, University of Connecticut, Storrs, Connecticut 06269-3046}
\author{Juha Javanainen}
\affiliation{Department of Physics, University of Connecticut, Storrs, Connecticut 06269-3046}

\begin{abstract}
Within the Bose-Hubbard model, we theoretically determine the stationary states of two distinguishable atoms in a one-dimensional optical lattice and compare with the case of two  identical bosons. A heterodimer has odd-parity dissociated states that do not depend on the interactions  between the atoms, and the lattice momenta of the two atomic species may have different averages even for a bound state of the dimer.  We discuss methods to detect the dimer. The different distributions of the quasimomenta of the two species may be observed in suitable time-of-flight experiments. Also, an asymmetry in the line shape as a function of the modulation frequency may reveal the presence of the odd-parity dissociated states when a heterodimer is dissociated by modulating the depth of the optical lattice.
\end{abstract}


\maketitle

\section{Introduction} \label{INTRODUCTION}

In an optical lattice the dispersion relation of atoms is different from free space, and both the motion of the atoms between the sites and the atom-atom interaction can be controlled experimentally.  Novel types of molecules with no free-space analogs are possible, for instance, a dimer bound by repulsive atom-atom interactions~\cite{WinklerK441}.  Several groups have elaborated on the theory~\cite{WinklerK441,Nygaard77, ValienteM41, Juha81, Sanders83}, and at least a qualitative agreement with the experiments~\cite{WinklerK441} has been established.  Most of the dimer models discuss two identical bosons, but theoretical analyses of two distinguishable atoms have been presented~\cite{Piil78, Petrosyan, Valiente81, Bruderer2010, Privitera2010}. There are also experiments on two different atomic species confined in an optical lattice~\cite{Ott2004, Ott2005, Gunter2006, Ospelkaus2006, Best2009, Catani2009, Lamporesi2010}, although at the moment we know of none that has specifically addressed lattice dimers.

Our initial contribution to the theory of the dimer of two identical bosons~\cite{Juha81} was based on the Bose-Hubbard model, and uniquely, solved for stationary states instead of Green's functions and scattering amplitudes. One can then straightforwardly analyze quantities such as the dissociation rate of a dimer induced by an external perturbation. More generally, we developed a detailed mathematical template to treat dimer problems in the one-dimensional Bose-Hubbard model~\cite{Juha81}. Our subsequent application of the template to the two-channel model of atom-atom interactions brought forward significant differences from the standard single-channel model~\cite{Sanders83}; see also~\cite{Nygaard77}.

The purpose of the present paper is to demonstrate how the template applies to a dimer of two distinguishable atoms. We discuss both the nagging factors of two and the structural changes in the theory that separate the cases of atoms with and without particle exchange symmetry. We also address two peculiar qualitative features of the heterodimer. First, even in the bound state the two atomic species may have different average quasimomenta, and the two species could seemingly drift apart. Our resolution is that the currents of both atomic species are  the same, and the atoms therefore move together. Nevertheless, the different distributions of quasimomenta are detectable in suitable time-of-flight experiments. In this context we also note the unusual property of a repulsively bound dimer, whether composed of identical or distinguishable atoms, that its effective mass is negative around zero center-of-mass momentum. As a result, the dimer will accelerate against a driving force. Second, there are twice as many dissociated states for the heterodimer than for a dimer of two identical atoms, and the additional continuum states may be chosen in such a way that they do not depend on the atom-atom interactions at all. The presence of these odd dissociated states may be detectable qualitatively as an asymmetry of the line shape  in an experiment in which the dimers are dissociated by modulating the depth of the optical lattice.

\section{Finite Lattice} \label{MODEL}

We consider a system of two distinguishable atoms, labelled $A$ and $B$, in a one-dimensional optical lattice. We assign the lattice an even number of lattice sites, $L$, that run from $k = 0, \ldots, L-1$, and take periodic boundary conditions such that $k = L$ is the same as $k = 0$.  Much of the emphasis in Ref.~\cite{Juha81} was on the correct way of taking the limit of an infinitely long lattice, $L \to \infty$.  Similar techniques apply here, but we will not discuss them anymore and the eventual limit $L \to \infty$ is implicitly assumed.

The Bose-Hubbard model for the Hamiltonian reads
\begin{align}
\frac{H}{\hbar} = & \, \sum_k \left[- \frac{J_A}{2} \left(a^\dagger_{k+1} a_k + a^\dagger_{k-1} a_k \right) + \frac{U_{AA}}{2} a^\dagger_k a^\dagger_k a_k a_k \right] \nonumber \\ & \, + \sum_k \left[- \frac{J_B}{2} \left(b^\dagger_{k+1} b_k + b^\dagger_{k-1} b_k \right) + \frac{U_{BB}}{2} b^\dagger_k b^\dagger_k b_k b_k \right] \nonumber \\ & \, + \sum_k \frac{U_{AB}}{2} a^\dagger_k b^\dagger_k a_k b_k \, .
\label{FIRSTHAM1}
\end{align}
Here $a_k$ and $b_k$ are the annihilation operators for the atomic species $A$ and $B$, $J_A$ and $J_B$ denote the respective site-to-site tunneling amplitudes, and $U_{AA}$, $U_{BB}$, and $U_{AB}$ characterize the strengths of the atom-atom interactions in each lattice site $k$.

Transforming from position representation to momentum representation with a discrete Fourier transformation of the operators as in~\cite{Juha81}, $a_k \to c_q$ and $b_k \to d_q$, puts the Hamiltonian in the form
\begin{align}
\frac{H}{\hbar} = & \, \sum_q \left[ \omega_A(q) c^\dagger_q c_q + \omega_B(q) d^\dagger_q d_q \right] \nonumber \\ & \, + \sum_{q_1,q_2,q_3,q_4} \!\!\! \delta_{q_1 + q_2, q_3 + q_4} \bigg{[}\frac{U_{AA}}{2L} c^\dagger_{q_1} c^\dagger_{q_2} c_{q_3} c_{q_4} \nonumber \\
& \, + \frac{U_{BB}}{2L} d^\dagger_{q_1} d^\dagger_{q_2} d_{q_3} d_{q_4} + \frac{U_{AB}}{2L} c^\dagger_{q_1} d^\dagger_{q_2} c_{q_3} d_{q_4} \bigg{]} \, ,
\label{FIRSTHAM2}
\end{align}
with
\begin{equation}
\omega_A(q) = - J_A \cos q \, ; \quad  \omega_B(q) = - J_B \cos q \, .
\label{OMEGAAOMEGAB}
\end{equation}
Unless otherwise noted, the lattice quasimomenta, or momenta for short, run over the first Brillouin zone $[-\pi,\pi)$ in steps of $2\pi/L$, and addition and comparison of two momenta are modulo $2\pi$.

The most general state vector for two different atoms is
\begin{equation}
\ket \psi = \sum_{p_1,p_2} A(p_1,p_2) c_{p_1}^{\dagger} d_{p_2}^{\dagger} \ket 0 \, ,
\label{STATE1}
\end{equation}
where $A(p_1, p_2)$ are expansion coefficients and $\ket 0$ is the particle vacuum. However, in order to separate the center-of-mass and internal motion of the dimer we resort to center-of-mass momentum, $P$, and relative momentum, $q$, such that $p_{1,2} = \half P \pm q$, and write
\begin{equation}
\ket \psi = \sum_{q} A(q)c_{\half P+q}^{\dagger} d_{\half P-q}^{\dagger} \ket 0 \, .
\label{STATE2}
\end{equation}
Here $\half P$ is the average lattice momentum per atom, taken to be in the first Brillouin zone, so that the nominal range of $P$ is from $-2\pi$ to $2\pi$.  The relative momentum $q$ still runs over the first Brillouin zone.  For identical bosons we have $A(q) = A(-q)$, but now there is no such exchange symmetry.  The vectors $c_{\half P + q}^{\dagger} d_{\half P-q}^{\dagger} \ket 0$ are a priori orthonormal for different $q$, so that we define the inner product between two states of the form of~\eq{STATE2} as
\begin{equation}
(A,B) = \sum_q A^*(q) B(q)\,,
\label{INNERPRODUCT}
\end{equation}
and use this inner product to normalize the states.  This inner product differs from the inner product we used for  indistinguishable bosons~\cite{Juha81} by a factor of two.

The separation of the center-of-mass motion succeeds to the extent that the Hamiltonian can be diagonalized separately for each $P$.  The time independent Schr\"odinger equation for the amplitudes can be cast as an equation for the expansion coefficients $A(q)$ in the form
\begin{equation}
- \Omega_P \cos(q + \phi) A(q) + \frac{U_{AB}}{2L} \sum_{q^\prime} A(q^\prime) = \frac{E}{\hbar} A(q) \, .
\label{TISE1}
\end{equation}
Since there is only one atom of each species, the $A$-$A$ and $B$-$B$ interactions are moot. Atom statistics is inoperative; the final results would be the same for two different species of fermions, and for a boson and a fermion. Finally, there is a degree of subtlety with the parameters $\Omega_P$ and $\phi$. Namely, we should have
\begin{equation}
J_A \cos (\half P +q) + J_B \cos (\half P -q) = \Omega_P \cos(q+\phi)\,.
\end{equation}
To achieve this, we amend the results from naive trigonometry so that they read
\begin{align}
\Omega_P = & \, \sqrt{J_A^2 + J_B^2 + 2 J_A J_B \cos(P)} \, , \label{OMEGAP} \\
\phi = & \, \arctan \left[ \frac{J_A - J_B}{J_A + J_B} \tan \left( \half P \right) \right] + \Phi(P) \, .
\label{PHI}
\end{align}
In Eq.~\eq{PHI} the function $\tan(\half P)$ diverges when $P$ approaches an odd multiple of $\pi$, and the branch of the $\arctan$ function normally used in numerics therefore jumps by $\pm \pi$ when $P$ crosses such a value.  The purpose of the added function $\Phi(P)$ is to correct for the abrupt jump, i.e., to select a proper branch of the $\arctan$ function.  $\Phi(P)$  is piecewise constant, has the value $0$ at $P=0$, and jumps by $\pm\pi$ at the divergences of $\tan(\half P)$ in such a way that $\phi$ remains a continuous function of $P$. Parameters akin to $\Omega_P$ and $\phi$ were also encountered in Refs.~\cite{Piil78, Valiente81}.

Let us next define the dimensionless quantities representing the energy of a state and the strength of the atom-atom interaction as follows,
\begin{equation}
\omega \equiv \frac{E}{\hbar \Omega_P} \, , \quad \Kappa \equiv \frac{U_{AB}}{2 \Omega_P} \, .
\label{DIMENSIONLESSQ}
\end{equation}
The equation for the eigenvalues of the dimensionless energy is then
\begin{equation}
f(\omega,\phi)=\frac{1}{L} \sum_q \frac{1}{\omega + \cos(q+\phi)} = \frac{1}{\Kappa} \, .
\label{EIGENVALUEEQ}
\end{equation}
We have  Eq.~\eq{EIGENVALUEEQ} both for indistinguishable and distinguishable bosons, but for identical bosons with the added rule that  $\phi\equiv0$. Given $\omega$, the coefficients $A(q)$ are obtained from
\begin{equation}
A(\omega, q) = \frac{C(\omega)}{\omega + \cos(q+\phi)} \, ,
\label{EIGENF}
\end{equation}
where unit normalization with respect to the inner product~\eq{INNERPRODUCT} is achieved with the choice
\begin{equation}
C(\omega) = \left[ \sum_q \frac{1}{\left[\omega + \cos \left(q + \phi \right) \right]^2} \right]^{-\half} \, .
\label{NORMALIZATIONC}
\end{equation}
\begin{figure}
\includegraphics[width=8.3cm]{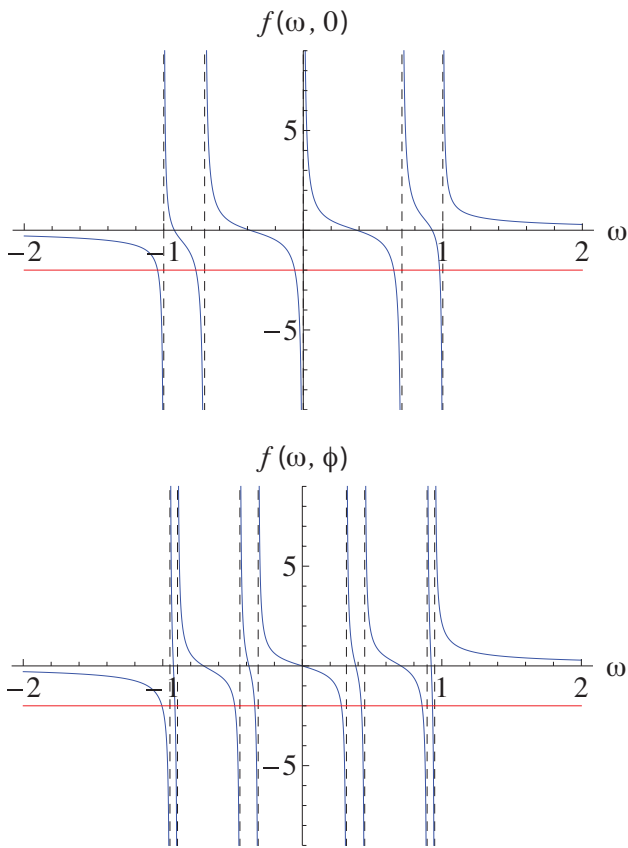}
\caption{(Color online) The plots of $f(\omega,0)$ (upper) and $f(\omega,\phi)$ with $J_A = 1$, $J_B = 1/3$, and $P=\half\pi$ (lower). The length of the lattice is $L=8$. Intersections of $f(\omega)$ and the horizontal lines at $f(\omega) = -2$ give the graphical solutions of $f(\omega)= 1/\Kappa$ for $\Kappa = - \half$. The dashed vertical lines are the asymptotes of $f(\omega)$.}
\label{FFIG}
\end{figure}

Compared to the case of identical particles, there are some structural changes in the theory. First, the definition of the parameter $\Kappa$ does not reduce to the one obtained for identical particles if one simply sets $J_A = J_B$. Second, Eq.~\eq{EIGENVALUEEQ} produces more solutions than was the case with identical atoms.

We illustrate the latter observation in Fig.~\ref{FFIG} with the plots of $f(\omega,\phi)$ in the case of two identical bosons~\cite{Juha81} (upper part) and in the present case (lower part). Here and below, to draw a figure we fix compatible units of energy and frequency, and express dimensional quantities as pure numbers with the units implied.  For two different species we have $J_A=1$, $J_B=1/3$, and $P=\half\pi$. 

Now consult the upper half of Fig.~\ref{FFIG}. Since $q$ and $-q$ are both included in the sum in Eq.~\eq{EIGENVALUEEQ}, except for $q=-\pi$, and since $\cos q=\cos(-q)$, for identical bosons with $\phi=0$ the function $f(\omega,0)$ as defined in Eq.~\eq{EIGENVALUEEQ} has $L/2+1$ asymptotes  in the interval $\omega\in[-1,1]$ over which its value switches from $+\infty$ to $-\infty$. A solution for the Schr\"odinger equation emerges whenever $f(\omega,0)=1/\Kappa$. There are therefore $L/2$ solutions to the identical-boson version of  Eq.~\eq{EIGENVALUEEQ} with $\omega\in(-1,1)$. Next move on to the lower half of Fig.~\ref{FFIG}. For two distinguishable atoms $\phi\ne0$ holds true, and there are $L$ asymptotes. This is because, as a matter of principle, the real number $\phi$ cannot be exactly equal to zero by accident, and no two values of $\cos(q+\phi)$ can be exactly the same for the given set of values of $q$. For indistinguishable atoms we therefore have $L-1$ solutions to Eq.~\eq{EIGENVALUEEQ} in the interval $\omega\in(-1,1)$.

In addition to these ``continuum'' states, there is also one solution with $|\omega|>1$, whether the bosons are identical or not. For attractive atom-atom interactions with $\Kappa<0$ this represents the usual bound dimer, with $\Kappa>0$ we have a repulsively bound dimer. The total number of solutions to Eq.~\eq{EIGENVALUEEQ} is therefore $L$ for distinguishable atoms, and $L/2+1$ for identical bosons. These numbers gratifyingly agree with the dimensions of the state space for the vectors~\eq{STATE2} and its counterpart for indistinguishable bosons~\cite{Juha81}.

\begin{figure}
\begin{center}
\includegraphics[width=8.3cm]{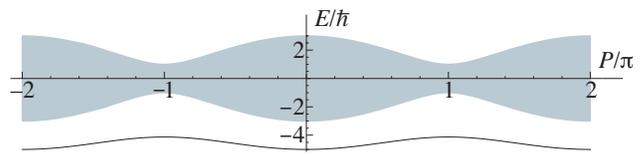}
\end{center}
\caption{(Color online) Illustration of the energy level structure for the relative motion of the atoms for varying center-of-mass momentum $P$. The grey band represents the dissociation continuum, and the curve at the bottom is the bound-state energy as a function of $P$. The parameters are $J_A=1$, $J_B=2$, and $U_{AB}=-8$, and the figure is for the continuum limit $L\rightarrow\infty$; see Sec.~\ref{CONTINUUMLIMIT}. Frequencies and energies are in arbitrary but consistent units.}
\label{BANDSTRUCTURE}
\end{figure}

In Fig.~\ref{BANDSTRUCTURE} we sketch the usual diagram~\cite{WinklerK441,Piil78} for  the energy level structure of the relative motion of the two atoms for varying center-of-mass momenta $P$. We set $J_A=1$, $J_B=2$, and $U_{AB}=-8$; in effect, the  tunneling frequency of atoms $A$, $J_A$, is chosen as the unit of frequency. For each value of $P$, the line at the bottom gives the corresponding bound-state energy and the grey band represents the range of the corresponding continuum states. Unlike for identical bosons, here the width of the dissociation continuum is not equal to zero for any center-of-mass momentum $P$.

\section{Continuum Limit} \label{CONTINUUMLIMIT}

We now focus on the limit $L \to \infty$ when the sum over the quasimomenta $q$ may be replaced by an integral,
\begin{equation}
\sum_q f(q) = \frac{L}{2\pi} \int_{-\pi}^\pi dq\, f(q) \, .
\label{CONTINUUMAPPROX}
\end{equation}
As before~\cite{Juha81}, even if we go to the limit $L\to \infty$ when the momentum $q$ becomes continuous, we imagine that all matrix elements, etc., are still calculated as discrete sums over $q$ in accordance with the inner product~\eq{INNERPRODUCT}, except that the sums are approximated as in Eq.~\eq{CONTINUUMAPPROX}. The factor $L/2\pi$ is therefore always retained with the integral over $q$, and it will manifest itself in the normalization coefficients of the state vectors.

\subsection{Bound State}

In the limit of a continuous $q$, the phase $\phi$ becomes irrelevant in the eigenvalue equation~\eq{EIGENVALUEEQ} and we have the same equation to solve as for identical particles.
The bound-state eigenvalue $\omega_b$, with $|\omega_b|>1$, is obtained directly by applying the continuum limit~\eq{CONTINUUMAPPROX} to Eq.~\eq{EIGENVALUEEQ}, and reads
\begin{equation}
\omega_b = {\rm sgn} (\Kappa) \sqrt{1 + \Kappa^2} \, .
\label{OMEGAB}
\end{equation}
The normalized bound state is
\begin{equation}
A_b(\omega_b,q) = \frac{|\Kappa|^{3/2}}{\sqrt{L|\omega_b|}}\frac{1}{\omega_b + \cos(q + \phi)}\,.
\label{NORMALISEDBOUNDS}
\end{equation}

The bound state affords an easy opportunity to gain some insights into the meaning of the angle $\phi$. To this end, we first note that the Heisenberg equation of motion of the atom number at the site $k$ immediately identifies the operator for the current from the site $k$ to the next site $k + 1$ for the species $A$ in the form
\begin{equation}
\hat{I}_{k \to k+1}^A = \half i J_A \left(a_{k+1}^\dagger a_k - a_{k}^\dagger a_{k+1} \right) \, ,
\label{CURRENTA}
\end{equation}
and likewise for species $B$. The expectation values of the currents in the state~\eq{STATE2} are
\begin{equation}
\langle \hat{I}_{k \to k+1}^{A,B} \rangle = \frac{J_{A,B}}{L}\sum_q |A(q)|^2 \sin(\half P \pm q) \, .
\label{CURRENTAEXP}
\end{equation}
In the bound state~\eq{NORMALISEDBOUNDS} and for the continuum limit~\eq{CONTINUUMAPPROX}, we have
\begin{equation}
\langle \hat{I}_{k \to k+1}^{A,B} \rangle_b = - \frac{J_{A,B}}{L \omega_b} \sin \left(\half P \mp \phi \right) \, .
\label{CURRENTAEXP2}
\end{equation}
The current is the same at every site, as it should be by virtue of the lattice translation symmetry of the state~\eq{STATE2}; in position representation the replacement $k \to k + 1$ changes the state by a global phase factor.

For clarity, let us temporarily take $\phi \simeq 0$, and $P \ll 1$. For a repulsively bound pair with $\omega_b > 1$, the current $\propto- P$ is then in the direction opposite to the center-of-mass momentum of the dimer. To understand this result, first note from Eq.~\eq{NORMALISEDBOUNDS} that for $\omega_b > 1$ the distribution of the relative momentum $q$ peaks around $\pm \pi$. As per Eq.~\eq{STATE2}, the momenta of the $A$ and $B$ atoms then tend to reside around $q=\half P \pm \pi$, so that, from Eq.~\eq{CURRENTAEXP} the currents are approximately proportional to $\sin(\half P \pm \pi) = -\sin(\half P)$. It seems likely to us that if a stationary dimer with $P = 0$ is put under a force $F$, we will have $\dot{P} \propto F$. The atoms, and the dimer, will then start accelerating where the currents go, against the force. As a result of the coupling of the external and internal degrees of freedom, a near-stationary repulsively bound dimer should have a negative effective mass. On the other hand, a dimer bound by attractive atom-atom interactions behaves normally on this score.

As another example consider an attractively bound pair, $\omega_b < -1$, with $\phi \neq 0$. In this case, the dominant lattice momenta for the $A$ and $B$ atoms come from the condition $q+\phi\simeq0$ in the form $\half P - \phi$ and $\half P + \phi$. This does not, however, mean that the $A$ and $B$ atoms have a tendency to separate from one another. In fact, using Eqs.~\eq{OMEGAP} and~\eq{PHI}, one sees immediately that the currents of the A and B atoms are equal,
\begin{equation}
\langle \hat{I}_{k \to k+1}^{A,B} \rangle_b = -\frac{J_A J_B \sin P}{L \omega_b \Omega_P} \, .
\label{CURRENTAEXP3}
\end{equation}
The motion of the atoms in the lattice is governed by two factors, the lattice momenta and the hopping matrix elements. The angle $\phi$ arranges itself in such a way that while the average lattice momenta of the two species are different, in a stationary state of the dimer the two species nevertheless have the same current and move together.

\subsection{Dissociated States}

To find the dissociated states, we proceed somewhat differently from Ref.~\cite{Juha81} and solve directly the continuum version of the Schr\"odinger equation. Specifically, we write
\begin{align}
\left[ \omega + \cos(q + \phi) \right] A(q) = \frac{\Kappa}{L} \sum_q A(q) = \frac{\Kappa}{2\pi} \int_{-\pi}^{\pi}dq\, A(q) \,.
\label{TISE2}
\end{align}
Now, defining $A_\phi(q) \equiv A (q-\phi)$ and noting that $A(q)$ may be taken to have the period $2\pi$, we have
\begin{equation}
\left( \omega + \cos q \right) A_{\phi}(q) = \frac{\Kappa}{2\pi} \int_{-\pi}^{\pi} dq\, A_{\phi}(q) \, .
\label{TISE3}
\end{equation}
By the symmetry of Eq.~\eq{TISE3}, we may always choose each solution to be either even or odd in the variable $q$. The even solutions $A_\phi^+(q)$ with $A_\phi^+(q) = A_\phi^+(-q)$ are almost the same as those we already found for identical bosons, where particle statistics allowed us to impose the condition $A(q) = A(-q)$ from the start. The  difference is that the normalization of the states is with respect to the inner product~\eq{INNERPRODUCT}, not the identical-boson version thereof.  Along the lines of Ref.~\cite{Juha81}, for the continuum states of the heterodimer with the energy $\omega_c$ the normalization condition is
\begin{align}
& \int_{-\pi}^{\pi} dq\, [A_c^\pm (\omega_c, q)]^*A_c^\pm (\omega_c, q) \nonumber \\
& =\sqrt{1-\omega_c^2}\, \left (\frac{2\pi}{L}\right )^2\!\delta(\omega_c - \omega_c^\prime) \, .
\label{NORMCONDITION}
\end{align}
The end result is an extra factor $\sqrt{2}$ in the state vectors. Undoing the transformation $A \to A_\phi$ we have the ``even'' (even with respect to $q = - \phi$) continuum states
\begin{align}
A_c^+(\omega_c, q) = & \, \frac{\sqrt{2}\, \Kappa \sqrt{1 - \omega_c^2}}{L\sqrt{\Kappa^2 + 1-\omega_c^2}} \bigg{\{}\mathbf{P} \frac{1}{\omega_c + \cos(q + \phi)}\nonumber \\
& \, + \frac{\pi\sqrt{1 - \omega_c^2}}{\Kappa}\,\delta[\omega_c + \cos(q+\phi)] \bigg{\}} \,,
\label{ECONTINUUMSTATE}
\end{align}
where $\mathbf{P}$ stands for the principal value integral.

The novelty of the heterodimer is the odd states with $A_\phi^-(q) = -A_\phi^-(-q)$, which would be the only permissible solutions for indistinguishable fermions. For the odd states Eq.~\eq{TISE3} reads
\begin{equation}
\left( \omega + \cos q \right) A_\phi^-(q) = 0 \, .
\label{TISE4}
\end{equation}
The odd continuum states do not depend at all on atom-atom interactions. They are clearly of the form
\begin{equation}
A_\phi^-(q) = A\,[\theta(q) - \theta(-q)] \delta(\omega_c + \cos q) \, ,
\label{OCONTINUUMSTATE}
\end{equation}
where $\theta$ is the unit step function and $A$ is a normalization constant.  After normalizing using~\eq{NORMCONDITION} and  reverting to the unshifted amplitudes, the ``odd'' continuum eigenstates are found to be
\begin{align}
A_c^-(\omega_c,q) = & \, \frac{\sqrt{2}\,\pi\sqrt{1 - \omega_c^2}}{L}\, [\theta(q+\phi)-\theta(-q-\phi)] \nonumber \\ & \, \times \delta \left[\omega_c + \cos(q + \phi) \right] \, .
\label{OCONTINUUMSTATE2}
\end{align}
The dissociated states~\eq{ECONTINUUMSTATE} and~\eq{OCONTINUUMSTATE2} come with the understanding that the continuous label $\omega_c \in (-1, 1)$ characterizing the energy is associated with the same density of states as identical bosons~\cite{Juha81}
\begin{equation}
\rho(\omega_c) = \frac{L}{2\pi} \frac{1}{\sqrt{1 - \omega_c^2}} \, ,
\label{DENSITYOFSTATES}
\end{equation}
except that here $\rho(\omega_c)$ applies separately for both even and odd states.

\section{Heterodimer Detection} \label{HETERODIMERDETECTION}

In Ref.~\cite{Juha81} we analyzed three different ways to detect a lattice dimer of identical bosons: measurement of the size of the bound state by detecting the occupation numbers (0, 1, or 2) of the lattice sites, study of quasimomentum distribution of the atoms, and detection of dimer dissociation by modulating the lattice depth in time. We now discuss these schemes in the case of a heterodimer of two distinguishable atoms.

\subsection{Pair correlations} 
Suppose there is precisely two distinguishable atoms in the lattice, and the number of atoms at each site is measured. The measurements are then repeated many times and the statistics compiled. Given the state  $\ket\omega$, the joint probability to find atom $A$ at site $k_1$ and atom $B$ at site $k_2$ is 
\bea
{\cal P}(k_1,k_2) & = & {\cal N} \mele{\omega}{ b^\dagger_{k_2}b_{k_2}a^\dagger_{k_1}a_{k_1}}{\omega} \nonumber 
\nopagebreak[1] \\  & = & \frac{\cal N}{L} |\alpha_{k_1-k_2}|^2 \, ,
\label{PROB}
\eea
where $\cal N$ is a normalization constant, and, just like before~\cite{Juha81}, 
\begin{equation}
\alpha_k = \frac{1}{\sqrt{L}}\sum_{k} A\left(\omega, q\right)e^{iqk}
\label{STATEP2}
\end{equation}
may be viewed as the wave function of the relative motion of the atoms in position (lattice site) representation.

As with identical bosons, for stationary states the probability ${\cal P}(k_1,k_2)$ only depends on the distance $k$ between the lattice sites. The variance of the distance between the detected atoms gives the size of the bound state,
\beq
(\Delta k)^2 =\frac{\sum_k k^2 |\alpha_k|^2}{\sum_k|\alpha_k|^2}   = \frac{1}{2\Kappa^2}\,.
\eeq 
The mathematics here is essentially the same as for identical bosons~\cite{Juha81}; the only difference lies in the factor of two in the definition of the parameter $\Kappa$.

\subsection{Brillouin zone mapping}
As discussed before~\cite{Ott2005,WinklerK441}, if the optical lattice is switched off on a time scale such that the structure of the state on a length scale of a single site has time to disappear adiabatically but the structure over the scale of the lattice as a whole remains, the quasimomentum distribution in the interval~$[\pi,\pi)$ will be converted into momentum distribution of the atoms released from the lattice. Ballistic expansion subsequently turns this distribution of momentum into a position distribution of the atoms.

Suppose that the quasimomentum distribution in the bound state of the dimer is measured in this way. The distributions for the two species $A$ and $B$ are predicted to be
\bea
{\cal M}_{A}(p)  &=& \mele{\psi}{c^\dagger_{p} c_{p}}{\psi} = \left| A_b \left( p - \half P \right)\right|,\\
{\cal M}_{B}(p)  &=& \mele{\psi}{d^\dagger_{p} d_{p}}{\psi} = \left| A_b \left(- p + \half P \right)\right|\,.
\eea
For two different species with $\phi\ne0$ these are in fact different. Normalized to unity in $p$, they read
\begin{align}
& f_{A,B}(p, P) = \frac{\abs{\Kappa(P)}^3}{2 \pi \sqrt{1 + \Kappa(P)^2}} \nonumber \\ & \, \, \times \frac{1}{\left\{ {\rm sgn} \left[ \Kappa(P) \right] \sqrt{1 + \Kappa(P)^2} + \cos \left(p - \half P \pm \phi \right) \right\}^2} \, ,
\label{MOMDIST}
\end{align}
with
\begin{equation}
\Kappa(P) = \frac{U_{AB}}{2 \Omega_P} \, .
\end{equation}

\begin{figure}
\begin{center}
\includegraphics[width=8.0cm]{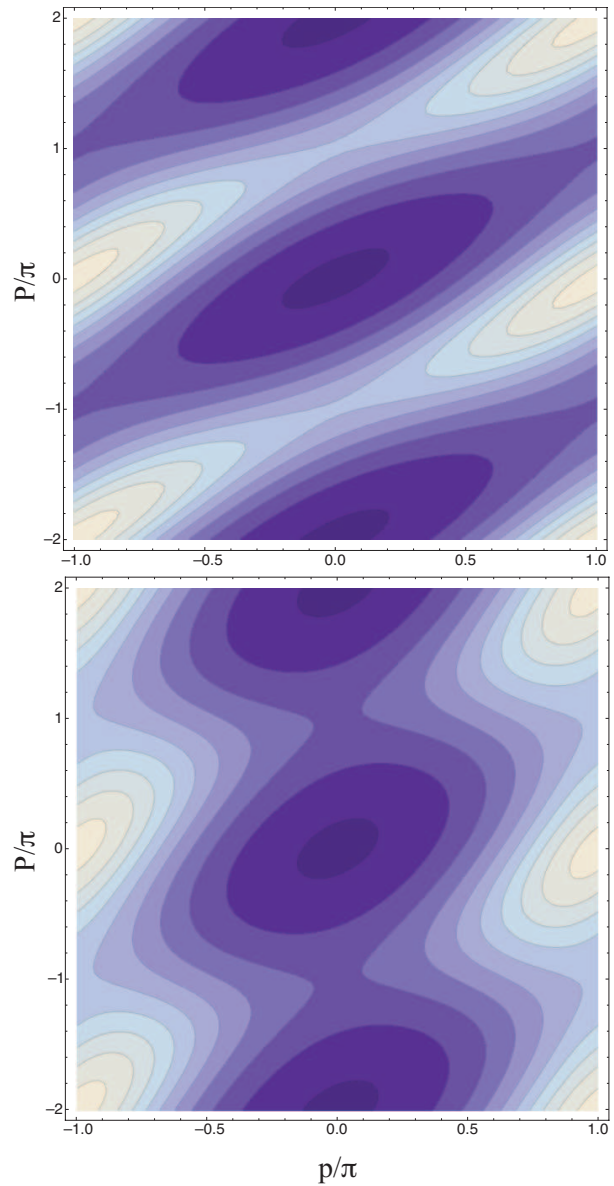}
\end{center}
\caption{(Color online) Contour plot of the momentum distribution of atomic species $A$ (top) and $B$ (bottom) in the bound state of the lattice dimer as a function of the quasimomentum of the detected atom $p$ and total momentum of the dimer $P$. The parameters are $J_A=1$, $J_B=2$, and $U_{AB}=16$. Lighter shading corresponds to larger value.}
\label{MOMDISTPLOTA}
\end{figure}

The momentum distribution for both atomic species in the repulsively bound state is depicted in Fig.~\ref{MOMDISTPLOTA} for the parameter values $J_A=1$, $J_B=2$, and $U_{AB}=16$. The value of $f_{A,B}$ is larger where the shading is lighter. Overall, the quasimomentum distribution is different for the two species. This does not lead to the dimer getting torn apart, but the difference could be detected directly in an experiment similar to what we have described here. The qualitative features of each distribution are similar to the case of identical bosons~\cite{WinklerK441,Juha81}, the most obvious difference being that for $J_A\ne J_B$ the distribution in the momentum $p$ of an individual atom is not flat for any center-of-mass momentum $P$ of the dimer. 

\subsection{Modulation spectroscopy}
Now suppose that the lattice is perturbed by periodically modulating the intensity of the lattice light in such a way that the tunneling rates  get modulated (approximately) sinusoidally, $J_A \rightarrow J_A + \Lambda_A \cos \nu t$ and $J_B \rightarrow J_B + \Lambda_B \cos \nu t$. Here $\nu$ denotes the the frequency of the intensity modulation, and $\Lambda_A$, $\Lambda_B$ are the respective modulation amplitudes for the atomic species $A$, $B$. It should be noted that, unlike with our convention in Ref.~\cite{Juha81},  $\Lambda_A$  and $\Lambda_B$ are dimensional quantities not scaled to the frequency $\Omega_P$. \commentout{In  general,  $\Lambda_A$ and $\Lambda_B$ need not be equal or proportional to the corresponding $J_{A}$ and $J_{B}$, but in a far-off resonant optical lattice both of these circumstances could apply if the $A$ and $B$ are hyperfine states of the same species. A more}A detailed analysis of the dependence of $J_A$ and $J_B$  on the lattice parameters, not to mention $\Lambda_A$  and $\Lambda_B$, is beyond our present scope.

The modulation makes a perturbation that couples the bound state of the dimer to the continuum states, and thereby affects the dissociation of the dimers. The corresponding Hamiltonian is
\begin{equation}
\frac{H'}{\hbar} = -\cos\nu t \,\sum_q \cos q\left(\Lambda_A\,c^\dagger_q c_q + \Lambda_B\,d^\dagger_q d_q \right)\,.
\label{PERTURBATION}
\end{equation}
This perturbation is diagonal in the center-of-mass momentum $P$. Its matrix elements between any two states of the form in Eq.\eq{STATE2} are given by
\begin{equation}
\mele{\psi_i}{\frac{H'}{\hbar}}{\psi_j} = - \Lambda_P M_{ij} \, \cos \nu t \, .
\end{equation}
Here
\begin{equation}
\Lambda_P = \sqrt{\Lambda_A^2 + \Lambda_B^2 + 2 \Lambda_A \Lambda_B \cos(P)} \,
\end{equation}
characterizes the strength of the modulation, and $M_{ij}$ is a dimensionless number that covers the structure of the dimer;
\begin{equation}
M_{ij} =  \frac{L}{2\pi} \int_{-\pi}^{\pi} dq \, \cos(q + \alpha) \, A^*_i(q) A_j(q) \, ,
\label{MELEINTEG}
\end{equation}
with
\begin{equation}
\alpha = \arctan \left[ \frac{\Lambda_A - \Lambda_B}{\Lambda_A + \Lambda_B} \tan(\half P) \right] + \Phi(P) \, ,
\label{ALPHA}
\end{equation}
where $\Phi(P)$ is the same specifier of the branch of the $\arctan$ function that was discussed after Eq.~\eq{PHI}.

Letting the subscripts $e$ and $o$ denote states in the continua of even and odd states, respectively, we have
\begin{align}
M_{be} = & \, M_{eb} = \left[ \frac{2|\Kappa|^3 (1 - \omega_e^2)}{L|\omega_b| (\omega_b^2 - \omega_e^2)} \right]^\half  \cos \beta \,, \label{MELEINTEGO1}\\
M_{bo} = & \, M_{ob} = -{\rm sgn}(\omega_b)\left[ \frac{2|\Kappa|^3(1 - \omega_o^2)}{L|\omega_b|(\omega_b - \omega_o)^2} \right]^\half \sin  \beta \,, \label{MELEINTEGO2}
\end{align}
where we have defined
\begin{equation}
\beta \equiv \alpha - \phi \,.
\end{equation}
Using the Golden Rule, we have the dissociation rates to the even and odd continua as
\begin{equation}
\Gamma_{j} = \frac{\pi |\Lambda_P|^2 |M_{bj}|^2}{2 \Omega_P} \rho(\Delta) \, .
\label{DISS}
\end{equation}
These dissociation rates are dimensional, not scaled to the parameter $\Omega_P$.
The argument of the density of states
\begin{equation}
\Delta = \omega_b \mp \frac{\nu}{\Omega_P} \,
\end{equation}
is the analog of detuning. It indicates the continuum state that is reached with conservation of energy from the bound state in a transition in which one ``quantum of energy'' of the size $\hbar\nu$ is either absorbed ($+$) or emitted ($-$) from the modulation of the lattice depth,
depending on whether the bound state is below or above the dissociation continuum. The dissociation rates of the dimer into the even and odd continua are
\begin{eqnarray}
\Gamma_{e} &=& \frac{|\Lambda_P|^2 |\Kappa|^3 \cos^2 \! \beta}{2 \Omega_P |\omega_b|}\frac{\sqrt{1 - \Delta^2}}{(\omega_b^2 - \Delta^2)},\label{DISSE}\\
\Gamma_{o} &=& \frac{|\Lambda_P|^2 |\Kappa|^3 \sin^2 \! \beta}{2 \Omega_P |\omega_b|} \frac{\sqrt{1 - \Delta^2}}{(\omega_b - \Delta)^2} \, .
\label{DISSO}
\end{eqnarray}

Potential dissociation of a dimer into the continuum of odd states is a feature not seen in the case of identical bosons~\cite{Juha81}. From Eqs.~\eq{DISSE} and~\eq{DISSO}, the relative strength of the dissociation rates is
\begin{equation}
\frac{\Gamma_{o}}{\Gamma_{e}} = \frac{\omega_b + \Delta}{\omega_b - \Delta} \tan^2 \! \beta \, .
\label{DISS3}
\end{equation}
If  the unperturbed tunneling rates $J_A$,  $J_B$ and the perturbations $\Lambda_A$, $\Lambda_B$ are equal for both species, as is likely in a far-off resonant lattice if the two species are Zeeman states or hyperfine states in the same atom, we have $\beta\simeq0$ and no dissociation to the odd channel. In this case, reverting everything to dimensional quantities and assuming otherwise identical parameter values, the dissociation rate to the even channel is the same as it would be for identical bosons if their atom-atom interaction parameter $U$ and the present interspecies interaction parameter $U_{AB}$ were related by  $U = U_{AB}/2$.

For strong interatomic interactions, $|\Kappa|\gg1$, we have $|\omega_b|\simeq|\Kappa| \gg |\Delta|$ for all allowable detunings $|\Delta|\le1$, so that $\omega_b^2-\Delta^2\simeq (\omega_b-\Delta)^2\simeq\Kappa^2$. The total dissociation rate $\Gamma_e + \Gamma_o$ would then be approximately the same as the dissociation rate to the even channel in the case $\beta=0$. However, dissociation into the odd channel is favored for modulation frequencies that attempt to break up a bound dimer to continuum states that are close in energy to the bound state, for instance, to $\Delta \sim -1$ for an attractively bound pair with $\omega_b<-1$. Dissociation into the odd channel, if present in the first place, makes the dimer loss rate an asymmetric function of the detuning $\Delta$, and might thus be observable experimentally especially for weak atom-atom interactions.

\section{Conclusion} \label{CONCLUSIONS}

Within the Bose-Hubbard model, we have determined the stationary states of two distinguishable atoms in a one-dimensional optical lattice. We have discussed, among other things, negative effective mass of a repulsively bound dimer, the varying roles of the quasimomentum distributions of the two species in the dimer, and dissociation of the heterodimer into a channel with no atom-atom interactions.  All of these aspects may have experimentally observable consequences.

\section*{Acknowledgments}

This work is supported in part by NSF Grant No. PHY-0967644.

\end{document}